\documentclass[%
 amsmath,amssymb,
 aps,
 prd,
twocolumn, superscriptaddress
]{revtex4-1}

\usepackage{float}
\usepackage[normalem]{ulem}

\usepackage{graphicx}
\usepackage{dcolumn}
\usepackage{bm}
\usepackage{hyperref}

\usepackage{xcolor}

\usepackage{color}

\begin{document}

\preprint{APS/123-QED} 

\title{ Anisotropic quark stars in $f(R)= R^{1+\epsilon}$ gravity }

\author{Juan M. Z. Pretel}
 \email{juanzarate@cbpf.br}
 \affiliation{
 Centro Brasileiro de Pesquisas F{\'i}sicas, Rua Dr. Xavier Sigaud, 150 URCA, Rio de Janeiro CEP 22290-180, RJ, Brazil
}

\author{S{\'e}rgio B. Duarte}
 \email{sbd@cbpf.br}
 \affiliation{
 Centro Brasileiro de Pesquisas F{\'i}sicas, Rua Dr. Xavier Sigaud, 150 URCA, Rio de Janeiro CEP 22290-180, RJ, Brazil
}

\date{\today}

\begin{abstract}
Within the metric formalism of $f(R)$ theories of gravity, where $R$ is the Ricci scalar, we study the hydrostatic equilibrium structure of compact stars with the inclusion of anisotropic pressure. In particular, we focus on the $f(R)= R^{1+\epsilon}$ model and we examine small deviations from General Relativity (GR) for $\vert \epsilon \vert \ll 1$. A suitable definition of mass function is explicitly formulated from the field equations and the value of the Ricci scalar at the center of each star is chosen such that it satisfies the asymptotic flatness requirement. We find that both the mass and the radius of a compact star are larger with respect to the general relativistic counterpart. Furthermore, we remark that the substantial changes due to anisotropy occur mainly in the high-central-density region.
\end{abstract}

\maketitle


\section{Introduction}

It is well known that the experimental tests of Einstein's general relativity at solar-system scales (i.e., the perihelion advance of Mercury, gravitational Doppler effect, light deflection, among others) have reached high precision \cite{Will2014}. Furthermore, GR has been tested in strong-gravity situations as the emission of gravitational waves emanating from a collision of compact objects \cite{Abbott221101, Abbott011102}. Consequently, standard GR has been the most successful and accepted theory of gravity to describe gravitational phenomena. Nevertheless, Einstein's theory has some limitations in the sense that a large portion of our observable Universe is not completely known. In fact, the interpretation of cosmic acceleration within the GR framework requires the introduction of dark energy --- an exotic form of matter with large negative pressure. In this direction, various modified theories of gravity have been proposed in the last decades, where the accelerated expansion of the Universe can arise from a modification of GR on large scales \cite{Starobinsky1980, Capozziello2002, Carroll2004, Hu2007, Nojiri2007, Amendola2007, Appleby2007, Odintsov2020}. Additional results on modified gravity models as alternatives to dark energy can be found in the review article by Koyama \cite{Koyama2016}. See also Refs.~\cite{PhysRevD.68.123512, Nojiri2008, Cognola2008, PhysRevD.103.044036, PhysRevD.103.124028} for a unified description of the inflationary era at early times with the dark energy epoch.

One of the simplest ways to modify GR is by replacing the scalar curvature $R$ in the conventional Einstein-Hilbert action by an arbitrary function of $R$, this is, the so-called $f(R)$ theories of gravity \cite{SotiriouFaraoni, Felice}. Such theories have been studied in order to explain the evolution of the early and present Universe, see e.g. Refs.~\cite{Capozziello2011, Nojiri2011, Clifton2012, Nojiri2017} for an extensive review. On the other hand, at astrophysical scale, the macroscopic properties of compact stars are also affected when the theory of gravity is modified. As a matter of fact, within the framework of $R^2$ gravity (i.e., the theory popularly known as the Starobinsky model \cite{Starobinsky1980}) and under a non-perturbative approach, the maximum mass of compact stars in the mass-radius diagram undergoes a noticeable increase due to the quadratic term \cite{Yazadjiev2014, Astashenok2015, Yazadjiev2015, Sbisa2020, Astashenok2020, Astashenok2021, AstashenokEPL, Nobleson2022, Jimenez2021}. It has been shown that the secondary component of the GW190814 event (reported by the LIGO-Virgo collaborations \cite{Abbott2020}) can be consistently described as a neutron star in the context of $R$-squared gravity \cite{Astashenok2020, Astashenok2021, AstashenokEPL}. Astashenok and Odintsov investigated the physical characteristics of non-rotating \cite{mnras214} and rotating \cite{mnras2630} neutron stars in $R^2$ gravity with the coupling of an axion scalar field. Additionally, for a comprehensive review on stellar structure models in the framework of modified theories of gravity formulated in both metric and metric-affine approaches, see Ref.~\cite{Olmo2020}.

B{\"o}hmer and collaborators \cite{Bohmer2008} showed that only small deviations from GR are needed to explain the flat galactic rotation curves. Namely, the gravitational Lagrangian in the form $f(R) \propto R^{1+ \epsilon}$ is a somewhat natural modification of Einstein gravity, where the dimensionless parameter $\epsilon$ is expressed in terms of the tangential velocity. In addition, this power-law $f(R)$ model has recently been considered for the explanation of the clustered galactic dark matter problem \cite{Sharma2020}, where the parameter $\epsilon$ is constrained to be $\mathcal{O}(10^{-6})$. The light deflection angle through the rotational velocity profile of typical nearby galaxies in the $R^{1+ \epsilon}$ gravitational background was calculated in Ref.~\cite{Sharma2021}. More studies on the physical implications of the power-law $f(R)$ theories of gravity can be found in Refs.~\cite{CapozzielloCT2006, CapozzielloCT2007, Martins2007, Jaryal2021, Sharma2022}.

On the other hand, it has been shown that the parameter $\epsilon$ within the context of $R^{1+\epsilon}$ gravity has a significant impact on the mass-radius diagrams of isotropic neutron stars \cite{Astashenok2020, Capozziello2015}. Although it is common to adopt isotropic perfect fluids to describe the compact star matter, there are well-supported reasons for the existence of anisotropy in superdense matter (see for instance Refs.~\cite{Chaichian2000, Ferrer2010, Horvat2011, Doneva2012, Silva2015, Yagi2015, Ivanov2017, Isayev2017, Biswas2019, Maurya2019, Pretel2020EPJC, Rahmansyah2020, Das2021, Das2021GRG, Deb2021APJ, Rahmansyah2021, Bordbar2022} and references therein for further discussions). In these works it has been shown that the presence of anisotropy allows to increase or decrease the maximum-mass values and, consequently, opens the possibility of obtaining more massive compact stars satisfying the astronomical observations.

Using a fully self-consistent non-perturbative approach within the Starobinsky model, the effect of anisotropies on the internal structure of neutron stars described by three different types of realistic equations of state (EoSs) has been studied by Folomeev \cite{Folomeev2018}. It was revealed that the inclusion of anisotropic pressure enables one to employ more stiff EoSs to model configurations that sufficiently satisfy the observational constraints. Moreover, anisotropic quark stars in $R$-squared gravity were investigated in Ref.~\cite{Panotopoulos2021}. More recently, Nashed \textit{et al.}~\cite{Nashed2021} studied anisotropic compact stars in the context of higher-order curvature theory of the $f(R)$ type. Malik and collaborators explored anisotropic compact spheres in the Starobinsky-like model \cite{Shamir2021, Malik2022a}, as well as in generalized modified gravity \cite{Shamir2019, Malik2022b, Malik2022c}. Using the Karmarkar condition and considering a logarithmic modification of the standard Starobinsky model, the gravitational collapse of an anisotropic system with heat flow was analyzed in Ref.~\cite{Usman2022}. See also Ref.~\cite{Ahmad2021} for a comparative analysis of self-consistent charged anisotropic spheres under embedded spacetime using the Karmarkar condition.

To the best of our knowledge, anisotropic compact stars have not yet been investigated in $R^{1+\epsilon}$ gravity model, so the purpose of the present study is to fill this gap. With this in mind, we will derive the stellar structure equations in such a gravity model and we will analyze the effect of anisotropy on compact stars. To do so, we will use the MIT bag model EoS for the radial pressure. It is worth emphasizing that quark matter within the MIT bag model framework was considered in the study of  astrophysical events by a large number of authors \cite{Brilenkov2013, Paulucci2014, Arbanil2016, Lugones2017, Deb2017Amsterdam, Chowdhury2020}. However, it is well known that the anisotropic aspects of EoS arise in the treatment of color superconductor states of dense quark matter (CFL and LOFF states) \cite{Kiriyama2006, Nasheeha2021}. It is out of the purpose of the present work to discuss such aspects, and we will consider the phenomenological ansatz proposed by Horvat \textit{et al.}~\cite{Horvat2011}. Our main goal here is to have an idea of the competition between gravitational effect with the $f(R)$ treatment and anisotropic aspect of the medium, on the quark star structure.

Our paper is organized as follows: In Sect.~\ref{Sec2} we briefly summarize metric $f(R)$ gravity, we present the field equations for a spherically symmetric system and introduce a mass function. Section \ref{Sec3} describes in detail the relativistic structure of anisotropic compact stars by means of the modified TOV equations. In the same section we present the EoS and the anisotropy profile. In Sect.~\ref{Sec4} we show our numerical results and we analyze the deviations of the physical quantities with respect to conventional GR. Finally, in Sect.~\ref{Sec5} we provide our conclusions. We will use the spacetime signature $(-,+,+,+)$ and geometric units with $G= 1 = c$. Nonetheless, our results will be given in physical units.


\section{Theoretical formalism}\label{Sec2}

\subsection{Metric $f(R)$ gravity}

Here we briefly summarize $f(R)$ theories of gravity within the metric formalism, where the conventional definitions of the curvature variables hold. Such theories are defined through the following action \cite{SotiriouFaraoni}
\begin{equation}\label{1}
    S = \frac{1}{16\pi}\int d^4x\sqrt{-g}f(R) + S_m ,
\end{equation}
where $R$ is the Ricci scalar, $g$ is the determinant of the metric tensor $g_{\mu\nu}$, and $S_m$ denotes the matter action. By varying the action with respect to the metric we obtain the field equations:
\begin{equation}\label{2}
f_R R_{\mu\nu} - \dfrac{1}{2}g_{\mu\nu}f - \nabla_\mu\nabla_\nu f_R + g_{\mu\nu}\square f_R = 8\pi T_{\mu\nu} ,
\end{equation}
with $T_{\mu\nu}$ being the matter energy-momentum tensor, $f_R \equiv df(R)/dR$, $\nabla_\mu$ stands for the covariant derivative associated with the Levi-Civita connection of the metric, and $\square \equiv \nabla_\mu\nabla^\mu$ is the d'Alembert operator. Besides, $R_{\mu\nu}$ represents the Ricci tensor and which is constructed solely from the connection. 

It can be noted that the field equations in $f(R)$ gravity are fourth-order differential equations in the metric since the Ricci scalar contains second-order derivatives, and it is evident that Eq.~(\ref{2}) reduces to the Einstein equation when $f(R)= R$. Inside a stellar fluid, in GR the Ricci scalar is defined by energy density and pressure, i.e. $R= -8\pi T$ where $T = g_{\mu\nu}T^{\mu\nu}$. However, in $f(R)$ gravity, both $g_{\mu\nu}$ and $R$ are dynamical fields, so that now the Ricci scalar is described by a second-order differential equation obtained by taking the trace of Eq.~(\ref{2}), namely
\begin{equation}\label{3}
    3\square f_R(R) + Rf_R(R) - 2f(R) = 8\pi T , 
\end{equation}
which means that $T=0$ no longer implies $R=0$ as in the pure general relativistic case. As we will see later, this indicates that non-linear functions in $R$ lead to a non-zero scalar curvature in the exterior region of a compact star.

\subsection{Field equations for a spherically symmetric system}

In order to examine the structure of compact stars in hydrostatic equilibrium, we consider a static and spherically symmetric system whose spacetime is described by the usual metric
\begin{equation}\label{4}
    ds^2 = -e^{2\psi}dt^2 + e^{2\lambda}dr^2 + r^2(d\theta^2 + \sin^2\theta d\phi^2) ,
\end{equation}
where $x^\mu = (t,r,\theta,\phi)$ are the Schwarzschild-like coordinates. The metric functions $\psi$ and $\lambda$ depend only on the radial coordinate $r$. 

The stellar matter distribution is assumed to be an anisotropic perfect fluid, i.e. it is described by the following energy-momentum tensor
\begin{equation}\label{5}
   T_{\mu\nu} = (\rho + p_t) u_\mu u_\nu + p_t g_{\mu\nu} - \sigma k_\mu k_\nu ,
\end{equation}
with $u^\mu$ being the four-velocity of the fluid and which satisfies the normalization condition $u_\mu u^\mu = -1$, $k^\mu$ is a unit radial four-vector so that $k_\mu k^\mu = 1$. Furthermore, $\rho$ is the energy density, $p_r$ the radial pressure, $p_t$ the tangential pressure and $\sigma \equiv p_t - p_r$ is the anisotropy factor. Accordingly, we can write $u^\mu = e^{-\psi}\delta_0^\mu$, $k^\mu= e^{-\lambda}\delta_1^\mu$ and the trace of the energy-momentum tensor (\ref{5}) takes the form $T= -\rho + 3p_r+ 2\sigma$. 

The four-divergence of expression (\ref{5}) provides the conservation law of energy and momentum as in conventional Einstein gravity, that is,
\begin{equation}\label{6}
    \nabla_\nu T_1^{\ \nu} = p_r' + (\rho + p_r)\psi' - \frac{2}{r}\sigma = 0 ,
\end{equation}
where the prime stands for differentiation with respect to the radial coordinate. In addition, $\square f_R$ is found to be 
\begin{align}
    \square f_R &= \frac{1}{\sqrt{-g}}\partial_\mu\left[ \sqrt{-g}\partial^\mu f_R \right]  \nonumber  \\
    &= \frac{1}{e^{2\lambda}}\left[ \left( \frac{2}{r} + \psi' - \lambda' \right)f_R' + f_R^{''} \right] .
\end{align}

Consequently, for the line element (\ref{4}) and energy-momentum tensor (\ref{5}), the non-zero components of the field equations (\ref{2}) are given by
\begin{widetext}
\begin{equation}\label{8}
    -\frac{f_R}{r^2} + \frac{f_R}{r^2}\frac{d}{dr}\left( re^{-2\lambda} \right) + \frac{1}{2}(Rf_R - f) + \frac{1}{e^{2\lambda}}\left[ \left( \frac{2}{r} - \lambda' \right)f_R' + f_R^{''} \right] = -8\pi\rho , 
\end{equation}
\vspace{-0.4cm}
\begin{equation}\label{9}
    -\frac{f_R}{r^2} + \frac{f_R}{e^{2\lambda}}\left( \frac{2\psi'}{r} + \frac{1}{r^2} \right) + \frac{1}{2}(Rf_R- f) + \frac{1}{e^{2\lambda}}\left( \frac{2}{r} + \psi' \right)f_R' = 8\pi p_r ,
\end{equation}
\vspace{-0.4cm}
\begin{equation}\label{10}
    \frac{f_R}{r^2}\left[ 1+ \frac{1}{e^{2\lambda}}(r\lambda' - r\psi' -1) \right] - \frac{1}{2}f + \frac{1}{e^{2\lambda}}\left[ \left( \frac{1}{r} + \psi' - \lambda' \right)f_R' + f_R^{''} \right] = 8\pi p_t ,
\end{equation}
\end{widetext}
and the dynamical equation for the scalar curvature (\ref{3}) becomes 
\begin{align}\label{11}
    \frac{3}{e^{2\lambda}}\left[ \left( \frac{2}{r} + \psi' - \lambda' \right)f_R' + f_R^{''} \right] =&\ 8\pi(-\rho + p_r + 2p_t)   \nonumber  \\
    &+ 2f - Rf_R .   
\end{align}

\subsection{Mass function}

It is convenient here to define a mass function for the stellar fluid. To do so, we will use the $00$-component of the field equations. In other words, the above Eq.~(\ref{8}) can be recast in the form 
\begin{align}\label{Mf1}
    \frac{d}{dr}\left( re^{-2\lambda} \right) &= 1- 8\pi r^2\rho - \left\lbrace (1- f_R)\frac{d}{dr}\left[ r(1- e^{-2\lambda}) \right]  \right.  \nonumber  \\
    &\left. \hspace{-1.0cm} + \frac{r^2}{2}(Rf_R- f) + \frac{r^2}{e^{2\lambda}}\left[ \left( \frac{2}{r}- \lambda' \right)f'_R+ f''_{R} \right]  \right\rbrace ,
\end{align}
this is, the metric function $\lambda$ is generated by the matter fields and by the terms related to the scalar curvature. Note that the scalar curvature is generated by the second-order differential equation (\ref{3}). Then, the integration of expression (\ref{Mf1}) leads to the following explicit result
\begin{equation}\label{Mf2}
    e^{-2\lambda} = 1 - \frac{2m}{r} ,
\end{equation}
where $m(r)$ plays the role of mass function and characterizes the mass enclosed within the radius $r$.  Thus, taking into account that $f_R' = R'f_{RR}$ and $f_R^{''} = R''f_{RR} + R'^2f_{RRR}$, such mass parameter can be written as
\begin{align}\label{Mf3}
    m =&\ 4\pi\int \rho r^2 dr + \frac{1}{2}\int\left\lbrace \frac{(1- f_R)}{r^2}\frac{d}{dr}\left[ r(1- e^{-2\lambda}) \right]  \right.  \nonumber  \\
    &\left. + \frac{1}{e^{2\lambda}}\left[ \left( \frac{2}{r}- \lambda' \right)R'f_{RR} + R''f_{RR}+ R'^2f_{RRR} \right] \right.  \nonumber  \\
    &\left. + \frac{1}{2}(Rf_R- f) \right\rbrace r^2dr . 
\end{align}

It is evident that when $f(R) = R$, the second integral vanishes and we recover the widely known expression in GR (where the mass becomes constant outside a star). Nevertheless, note that here the scenario is different from Einstein gravity, even in the outer region of a compact star where $\rho = 0$, the expression (\ref{Mf3}) generates an extra mass contribution due to the Ricci scalar. For the specific function $f(R)= R + \alpha R^2$, our generalized version (\ref{Mf3}) reduces to the expression given in Ref.~\cite{Jimenez2021}.


\section{ Modified TOV equations in $f(R)$ gravity }\label{Sec3}

Here we will adopt a non-perturbative approach where one looks for solutions of the exact fourth-order differential equations with respect to the metric functions. In order to construct static anisotropic compact stars, we need to derive a modified version of the TOV equations in $f(R)$ gravity. To do so, we are going to properly combine the field equations (\ref{8})-(\ref{9}) together with Eqs.~(\ref{6}) and (\ref{11}). From Eqs.~(\ref{8}), (\ref{9}) and (\ref{11}), one can obtain the following expressions, respectively
\begin{widetext}
\begin{equation}\label{15}
    \lambda' = \frac{1}{r(2f_R + rR'f_{RR})}\left[ 8\pi\rho r^2e^{2\lambda} + \frac{e^{2\lambda}}{2}(r^2Rf_R - r^2f - 2f_R) + f_R + 2rR'f_{RR} + r^2(R''f_{RR} + R'^2f_{RRR})  \right] ,
\end{equation}
\vspace{-0.4cm}
\begin{equation}\label{16}
    \psi' = \frac{1}{2r(2f_R + rR'f_{RR})}\left[ r^2e^{2\lambda}\left( 16\pi p_r+ f- Rf_R \right) + 2f_R\left( e^{2\lambda}- 1 \right) - 4rR'f_{RR} \right] ,
\end{equation}
\vspace{-0.4cm}
\begin{equation}\label{17}
    R'' = \frac{1}{3f_{RR}}\left\lbrace e^{2\lambda}\left[ 8\pi(-\rho + p_r + 2p_t) + 2f - Rf_R \right] - 3R'^2f_{RRR} \right\rbrace + \left( \lambda' - \psi' - \frac{2}{r} \right)R' .
\end{equation}

In view of Eqs.~(\ref{16}) and (\ref{17}), Eq.~(\ref{15}) becomes 
\begin{align}\label{18}
    \lambda' =& \ \frac{1}{2rf_R}\left\lbrace f_R(1 - e^{2\lambda}) + \frac{r^2e^{2\lambda}}{3}\left[ 8\pi(2\rho+ p_r+ 2p_t) + \frac{1}{2}(Rf_R + f) \right] \right\rbrace  \nonumber  \\
    &- \frac{R'f_{RR}}{2f_R(2f_R + rR'f_{RR})}\left[ \frac{r^2e^{2\lambda}}{2}(16\pi p_r + f - Rf_R) - f_R(1- e^{2\lambda}) - 2rR'f_{RR} \right] ,
\end{align}
or alternatively, 
\begin{align}\label{19}
    \lambda' =&\ \frac{1}{2r(2f_R + rR'f_{RR})}\left\lbrace 2f_R(1- e^{2\lambda}) + \frac{r^2e^{2\lambda}}{3}\left[ 16\pi(2\rho+ p_r + 2p_t) +Rf_R + f \right] \right.  \nonumber  \\
    &\left. + \frac{rR'f_{RR}}{f_R}\left[ 2f_R(1 - e^{2\lambda}) + \frac{r^2e^{2\lambda}}{3}( 16\pi(\rho - p_r + p_t) + 2Rf_R -f ) + 2rR'f_{RR} \right]  \right\rbrace .
\end{align}

Therefore, within the context of $f(R)$ modified theories of gravity, the relativistic structure of a static compact star in the presence of anisotropic pressure is described by Eqs.~(\ref{16}), (\ref{17}), (\ref{19}) and (\ref{6}), rewritten as 
\begin{align}
    \frac{d\psi}{dr} &= \frac{1}{2r(2f_R+ rR'f_{RR})}\left[ r^2e^{2\lambda}(16\pi p_r + f - Rf_R) + 2f_R\left( e^{2\lambda}-1 \right) - 4rR'f_{RR} \right] ,  \label{TOV1}  \\
    \frac{d\lambda}{dr} &= \frac{1}{2r(2f_R+ rR'f_{RR})}\left\lbrace 2f_R\left(1- e^{2\lambda}\right) + \frac{r^2e^{2\lambda}}{3}\left[ 16\pi(2\rho + 3p_r + 2\sigma) + Rf_R +f \right] \right.  \nonumber  \\
    &\hspace{2cm} \left. +\frac{rR'f_{RR}}{f_R}\left[ 2f_R\left(1- e^{2\lambda}\right) + \frac{r^2e^{2\lambda}}{3}\left( 16\pi\rho + 16\pi\sigma + 2Rf_R - f \right) + 2rR'f_{RR} \right] \right\rbrace  ,  \label{TOV2}   \\
    \frac{d^2R}{dr^2} &= \frac{1}{3f_{RR}}\bigg\lbrace e^{2\lambda}\left[ 8\pi(-\rho + 3p_r+ 2\sigma) + 2f -Rf_R \right] -3R'^2f_{RRR} \bigg\rbrace + \left( \lambda' - \psi'- \frac{2}{r} \right)R' ,  \label{TOV3}  \\
    \frac{dp_r}{dr} &= -(\rho + p_r)\psi' + \frac{2}{r}\sigma ,  \label{TOV4}
\end{align}
\end{widetext}
where the particular case $\sigma= 0$ corresponds to TOV equations describing isotropic compact stars in $f(R)$ gravity \cite{Yazadjiev2015}. Equation (\ref{TOV3}) will play a crucial role in the radial behavior of the scalar curvature both inside and outside the star. In Einstein gravity, this equation is reduced to $R = 8\pi(\rho - 3p_r - 2\sigma)$, so that in the exterior region of the star we have $R=0$ and hence the Schwarzschild solution is valid. 

We can notice that the above system of equations (\ref{TOV1})-(\ref{TOV4}) correspond to three first-order and one second-order ordinary differential equations, and which contains a set of six variables $\psi$, $\lambda$, $R$, $\rho$, $p_r$ and $\sigma$ to be determined. Thus, given a barotropic EoS for radial pressure in the form $p_r= p_r(\rho)$ and an anisotropy relation for $\sigma$, only five boundary conditions are required to solve such a system inside the star. Indeed, by ensuring regularity of the geometry at the center of the star, we establish the boundary conditions
\begin{align}\label{BC1}
    \rho(0) &= \rho_c ,   &   \psi(0) &= \psi_c ,   &    \lambda(0) &= 0 ,  \nonumber   \\
    R(0) &= R_c ,   &   R'(0) &= 0 ,
\end{align}
where $\rho_c$ and $R_c$ are the values of the central energy density and central scalar curvature, respectively. In the meanwhile, outside the star, the solution is defined by Eqs.~(\ref{TOV1})-(\ref{TOV3}), where the energy density and pressures vanish ($\rho =p_r= p_t =0$) and hence the EoS is not needed any longer. The surface of the star is found when the radial pressure vanishes, i.e. $p_r(r_{\rm sur})= 0$. It is therefore convenient to settle the following junction conditions at the stellar surface 
\begin{align}\label{BC2}
    \psi_{in} (r_{\rm sur}) &= \psi_{out} (r_{\rm sur}) ,   &   \lambda_{in} (r_{\rm sur}) &= \lambda_{out} (r_{\rm sur}) ,  \nonumber  \\
    R_{in} (r_{\rm sur}) &= R_{out} (r_{\rm sur}) ,   &   R'_{in} (r_{\rm sur}) &= R'_{out} (r_{\rm sur}) .
\end{align}

Additionally, constrains on the Ricci scalar and mass function come from the asymptotic flatness requirement
\begin{align}\label{BC3}
    \lim_{r \rightarrow \infty} R(r) &= 0 ,   &   \lim_{r \rightarrow \infty} m(r) = \rm constant ,
\end{align}
namely, $R_c$ must be chosen so that it satisfies the requirement (\ref{BC3}) at infinity. In turn, the central value of the metric function $\psi$ in Eq.~(\ref{BC1}) is fixed by requiring that the spacetime geometry be asymptotically flat, i.e. $\psi(r\rightarrow \infty) \rightarrow 0$. Consequently, by bearing in mind Eq.~(\ref{Mf2}), the total gravitational mass of the star $M$ will be determined from the asymptotic behavior
\begin{equation}\label{27}
    M \equiv \lim_{r \rightarrow \infty} \frac{r}{2}\left( 1- \frac{1}{e^{2\lambda}} \right) .
\end{equation}

\subsection{ $f(R) = R^{1+\epsilon}$ gravity }

The solution of the system of equations (\ref{TOV1})-(\ref{TOV4}) with boundary conditions (\ref{BC1}) and (\ref{BC2}), characterizes completely the static background of an anisotropic compact star within the metric $f(R)$ formalism. In fact, to obtain such a solution, one has to specify the particular model of $f(R)$ gravity. An interesting class of models are the power-law models given by $f(R) \sim R^n$ (where $n \in \mathbb{R}$) because they are related to the existence of Noether symmetries \cite{Capozziello2007}. Capozziello and collaborators \cite{CapozzielloCT2007} showed that such gravity model may represent a good candidate to solve both the dark energy problem at cosmological level and the dark matter one at galactic scale with the same value of the slope $n$ of the higher-order gravitational theory. Following Ref.~\cite{Astashenok2020}, we assume that $n = 1+\epsilon$ so that we can study small deviations with respect to GR for $\vert\epsilon \vert \ll 1$. It is worth noting that this type of corrections emerges in one-loop regularization and renormalization process in curved spacetime. In this perspective, we can write $f(R)$ as a first-order Taylor expansion
\begin{equation}\label{28}
    R^{1+\epsilon} \simeq R+ \epsilon R\ln R ,
\end{equation}
this is, the correction term to the standard Einstein-Hilbert action is logarithmic and for $\epsilon =0$ we retrieve the pure general relativistic case. Therefore, $f_R = 1+ \epsilon+ \epsilon\ln R$, $f_{RR}= \epsilon/R$, $f_{RRR}= -\epsilon/R^2$ and which will be replaced into the modified TOV equations (\ref{TOV1})-(\ref{TOV4}) in order to determine the metric functions and thermodynamic quantities of an anisotropic compact star. In the present work we will use typical values for the leading parameter $\epsilon$ as in Refs.~\cite{Astashenok2020, Capozziello2015}. In fact, it has been argued that the results of physical interest are obtained for $\epsilon <0$ \cite{Capozziello2015}.

\subsection{ Equation of state and anisotropy profile }

Similar to the construction of anisotropic compact stars in Einstein gravity, to close the system of equations (\ref{TOV1})-(\ref{TOV4}) one needs to specify an EoS (this is, the microphysical relation between radial pressure and energy density by means of equation $p_r = p_r(\rho)$) and also assign an anisotropy function $\sigma$ since there is now an extra degree of freedom $p_t$. To explore quark stars in $R^{1+\epsilon}$ gravity, we employ the MIT bag model EoS for the dense matter involved, given by 
\begin{equation}\label{MITbagEoS}
    p_r = b(\rho - 4B) ,
\end{equation}
which describes a self-gravitating fluid composed by up, down, and strange quarks. The constant $b$ usually varies from 0.28 to $1/3$, and the bag constant $B$ lies in the range $0.982B_0< B <1.525B_0$ where $B_0 = 60\ \rm MeV/fm^3$ \cite{Paschalidis2017} In our study, we will consider the particular case $b= 1/3$ and $B= B_0$. Nevertheless, we must point out that values for the bag constant can be consistently determined together with other parameters in order to describe some compact objects observed in nature \cite{Delgado2022}.

We will consider values for the central energy density in the range $\rho_c \in \left[ 0.5, 4.0 \right] \times 10^{15}\, \rm g/cm^3$, which are typical values used in GR \cite{Arbanil2016, Rodrigues2011}. Notice that such a range of central densities is larger than the nuclear saturation density, i.e., $\rho/\rho_0> 1$, where $\rho_0 = 2.8 \times 10^{14}\, \rm g/cm^3$.

In addition to the EoS for radial pressure, we will use the anisotropy ansatz suggested by Horvat and collaborators \cite{Horvat2011} to model anisotropic matter inside compact stars, namely
\begin{equation}\label{AnisoProfile}
    \sigma = \beta p_r\mu ,
\end{equation}
or alternatively,
\begin{equation}
     p_t = p_r\left[ 1+ \beta(1- e^{-2\lambda}) \right], 
\end{equation}
with $\mu(r) \equiv 2m/r$ being the compactness of the star. In the non-relativistic limit, when the pressure contribution to the energy density is negligible, the effect of anisotropy vanishes in the hydrostatic equilibrium equation. This is in good agreement with the assumption that anisotropy may arise only at high densities of matter. Another advantage of such profile is that the stellar fluid becomes isotropic at the origin since $\mu \sim r^2$ when $r \rightarrow 0$. It is also commonly known as quasi-local ansatz in the literature where $\beta$ measures the degree of anisotropy inside the star and in principle can assume positive or negative values \cite{Horvat2011, Doneva2012, Silva2015, Yagi2015, Folomeev2018, Pretel2020EPJC, Rahmansyah2020, Rahmansyah2021}. Here we will consider values of $\beta$ for which appreciable changes in the mass-radius diagrams can be visualized.


\section{Numerical results and discussion}\label{Sec4}

\begin{figure}
  \includegraphics[width=8.4cm]{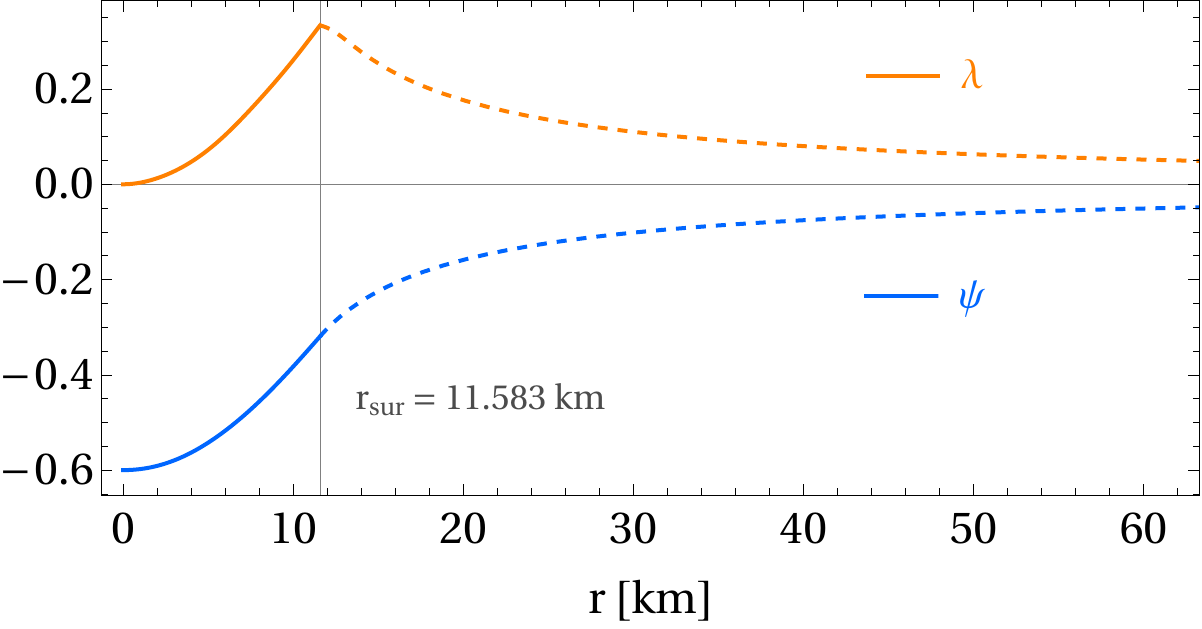}
  \includegraphics[width=8.65cm]{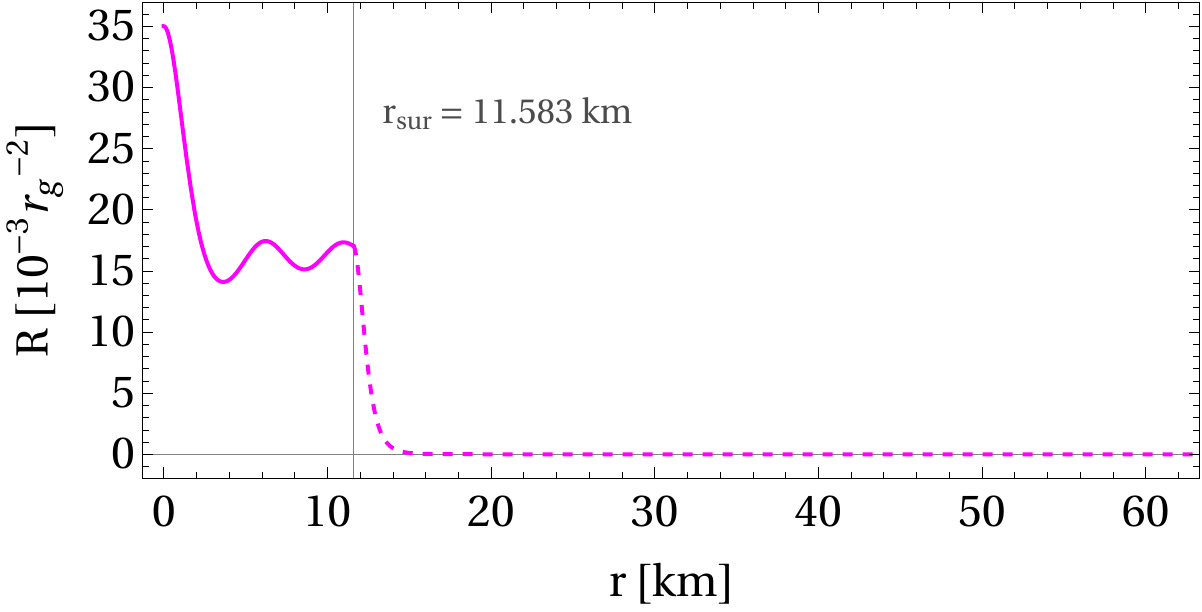}
  \caption{\label{figure1} Numerical solution of the set of modified TOV equations (\ref{TOV1})-(\ref{TOV4}) for a given central density $\rho_c = 1.0 \times 10^{18} \rm kg/m^3$ with MIT bag model EoS (\ref{MITbagEoS}) and anisotropy profile (\ref{AnisoProfile}) in $R^{1+\epsilon}$ gravity model, where we considered $\epsilon = -0.002$ and $\beta = 0.4$. The metric functions (upper panel) and Ricci scalar (lower panel) are displayed as functions of the radial coordinate. In each plot, the solid and dotted lines correspond to the interior and exterior solutions, respectively. It can be observed that the Ricci scalar goes to zero as we move away from the stellar surface according to the asymptotic flatness requirement (\ref{BC3}). }  
\end{figure}

Using the boundary conditions (\ref{BC1}) and (\ref{BC2}) together with the first-order expansion (\ref{28}), we numerically solve the modified TOV equations (\ref{TOV1})-(\ref{TOV4}) inside and outside the star with EoS (\ref{MITbagEoS}) and anisotropy profile (\ref{AnisoProfile}). Here we have to specify the numerical values of the free parameters $\epsilon$ and $\beta$. For the interior problem, this is carried out by integrating from the center of the star to the surface where the radial pressure vanishes, and for the exterior problem, by fulfilling the asymptotic flatness requirement (\ref{BC3}). In other words, the central value of the Ricci scalar is chosen so that asymptotically $R(r\rightarrow\infty) \rightarrow 0$ and hence $R_c$ can only assume a unique value for each stellar configuration. In particular, for a central energy density $\rho_c = 1.0 \times 10^{18}\ \rm{kg}/\rm{m}^3$, we display in Fig.~\ref{figure1} the interior and exterior solution for the metric functions and Ricci scalar in $f(R)= R^{1+ \epsilon}$ gravity. Analogously to the conventional GR theory, Minkowski spacetime is asymptotically required, i.e. $\psi$ and $\lambda$ approach zero at large distances. From the lower panel, we observe that there exist a significant contribution to the scalar curvature immediately after the surface and it decreases monotonically with $r$. This means that the Schwarzschild metric is not adequate to describe the exterior spacetime of a compact star in $f(R)$ gravity, it could only be a well-behaved limit at a sufficient distance from the surface.

\begin{figure*}
  \includegraphics[width=8.6cm]{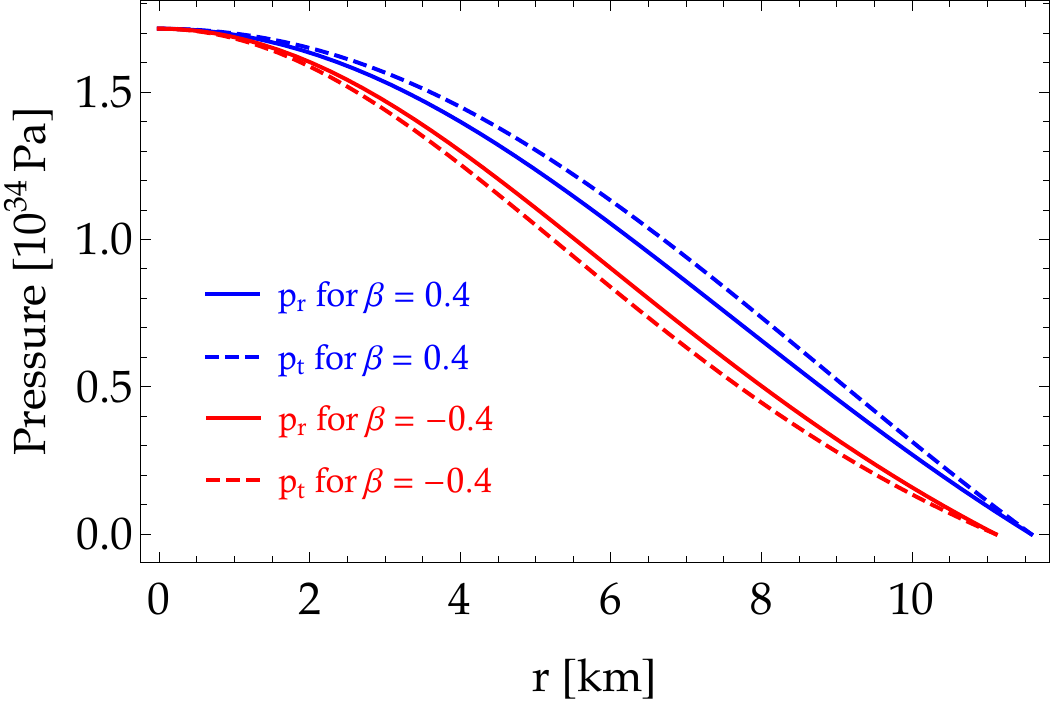}
  \includegraphics[width=8.6cm]{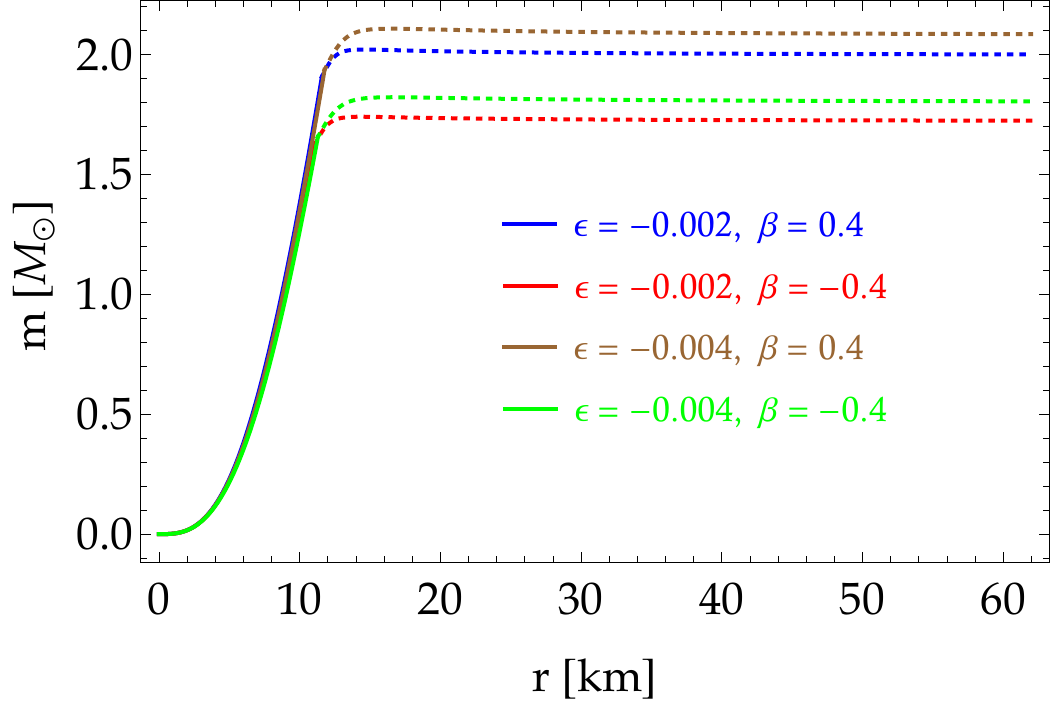}
  \caption{\label{figure2} Left panel: Radial (solid lines) and tangential (dashed lines) pressure as a function of the radial coordinate for anisotropic quark stars within the framework of $f(R) = R^{1+\epsilon}$ gravity with $\epsilon = -0.002$ and two specific values of $\beta$. For each configuration the stellar surface is determined when the radial pressure vanishes. Right panel: Radial behavior of the mass function inside (solid lines) and outside (dotted lines) a quark star for different values of the parameters $\epsilon$ and $\beta$. The graphs are plotted for the MIT bag model EoS and central density $\rho_c = 1.0 \times 10^{18} \rm kg/m^3$. }  
\end{figure*}

For the central energy density considered above, the radial and tangential pressures are shown in the left panel of Fig.~\ref{figure2} for $\epsilon = -0.002$ and two specific values of $\beta$. As expected, the fluid is isotropic at the stellar origin and this ensures regularity. In turn, the radius of the quark star decreases (increases) for negative (positive) values of $\beta$ as compared to the isotropic case (see also Table \ref{table1}). Furthermore, the mass function is displayed in the right panel of the same figure. In Einstein gravity the gravitational mass is given by the first integral of Eq.~(\ref{Mf3}), it grows from the center to the surface and always assumes a constant value in the exterior region of the star. Nevertheless, in the $R^{1+\epsilon}$ gravity model, the logarithmic term in Eq.~(\ref{28}) generates an extra mass contribution through the second integral of Eq.~(\ref{Mf3}). As a result, the total mass at infinity increases with respect to the mass measured at the surface $m(r_{\rm sur})$. This can be clearly observed in the data recorded in Table \ref{table1}, and thus our results are in very good agreement with those reported in Ref.~\cite{Astashenok2020}. In the literature it is common to interpret such additional mass as an effective mass due to a ``gravitational sphere'' outside the star \cite{Astashenok2015}. Namely, one finds that with the emergence of this sphere, the total gravitational mass (\ref{27}) increases as $\epsilon$ is more negative, compared to the pure GR case. From the same plot, note further that the role of the anisotropic pressure is to increase the gravitational mass as $\beta$ increases.

\begin{figure}
  \includegraphics[width=8.6cm]{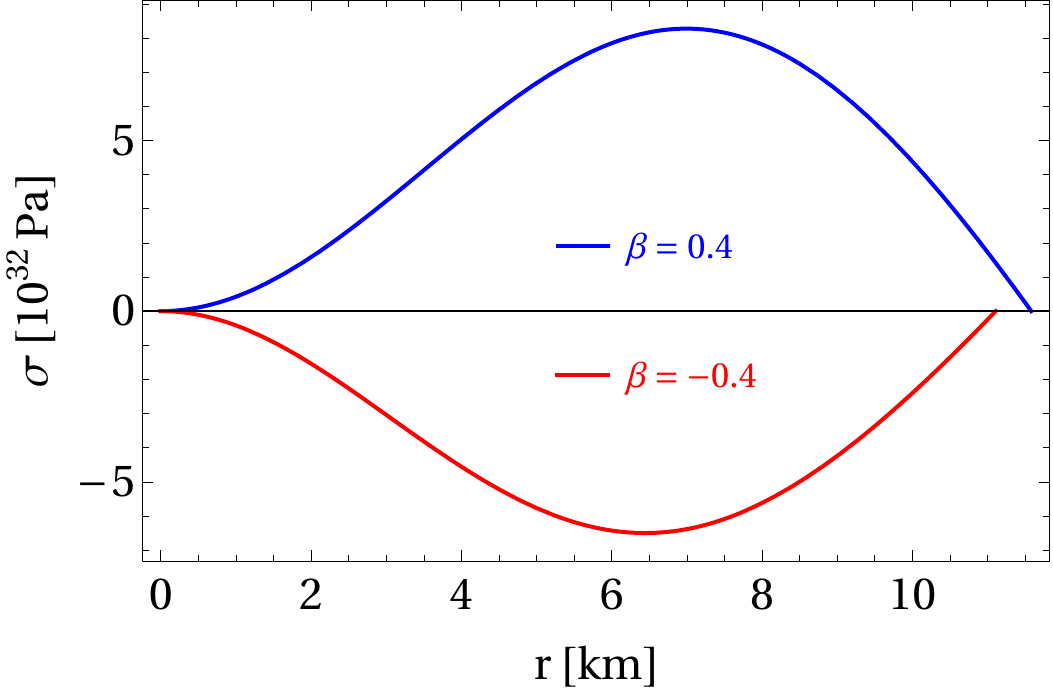}
  \caption{\label{figure3} Radial behavior of the anisotropy factor for a quark star with central density $\rho_c = 1.0 \times 10^{18} \rm kg/m^3$. As in Fig.~\ref{figure2}, we have considered $\epsilon = -0.002$ and two values of $\beta$. }  
\end{figure}

Figure \ref{figure3} displays the radial behavior of the anisotropy factor for a given central density. It is more pronounced in the intermediate regions of the star and can be positive (negative) if $\beta$ is positive (negative). As predicted by Eq.~(\ref{AnisoProfile}), the anisotropy vanishes both at the center and at the surface of the star. Following Refs.~\citep{Shamir2021, Hernandez2004, Pretel2019, Bogadi2021}, the second term on the right-hand side of Eq.~(\ref{TOV4}) can be interpreted as a force due to anisotropy. Namely, this force is directed outward when $2(p_t-p_r)/r > 0$ and inward if $p_t < p_r$.

\begin{figure*}
  \includegraphics[width=8.6cm]{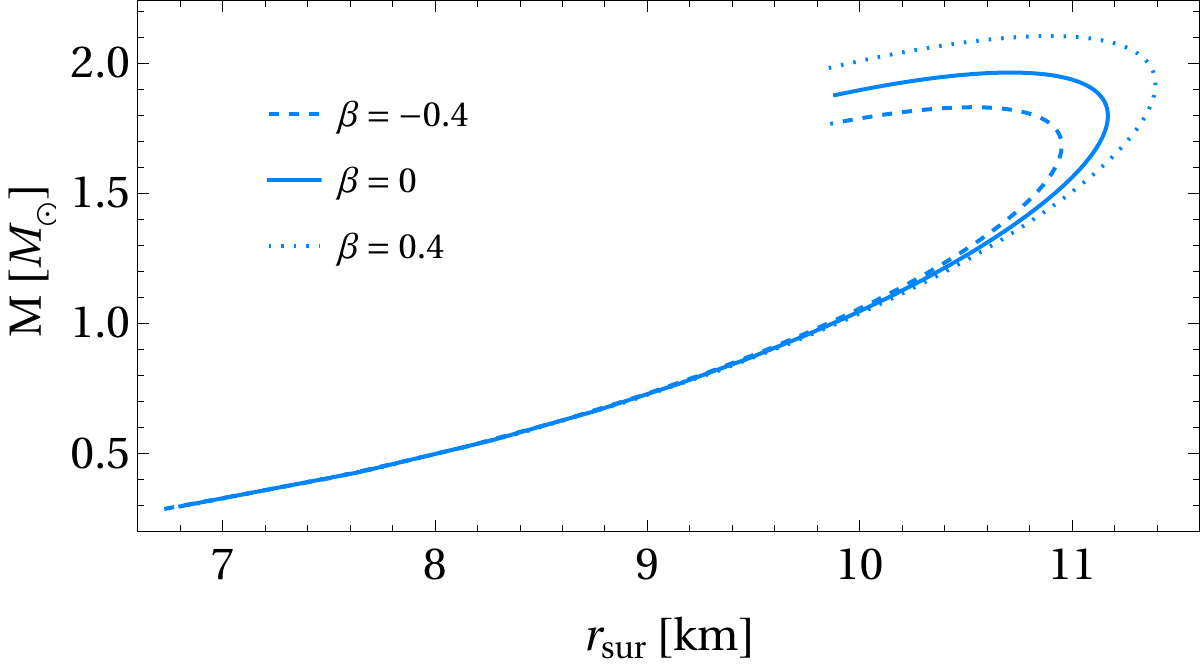}
  \includegraphics[width=8.6cm]{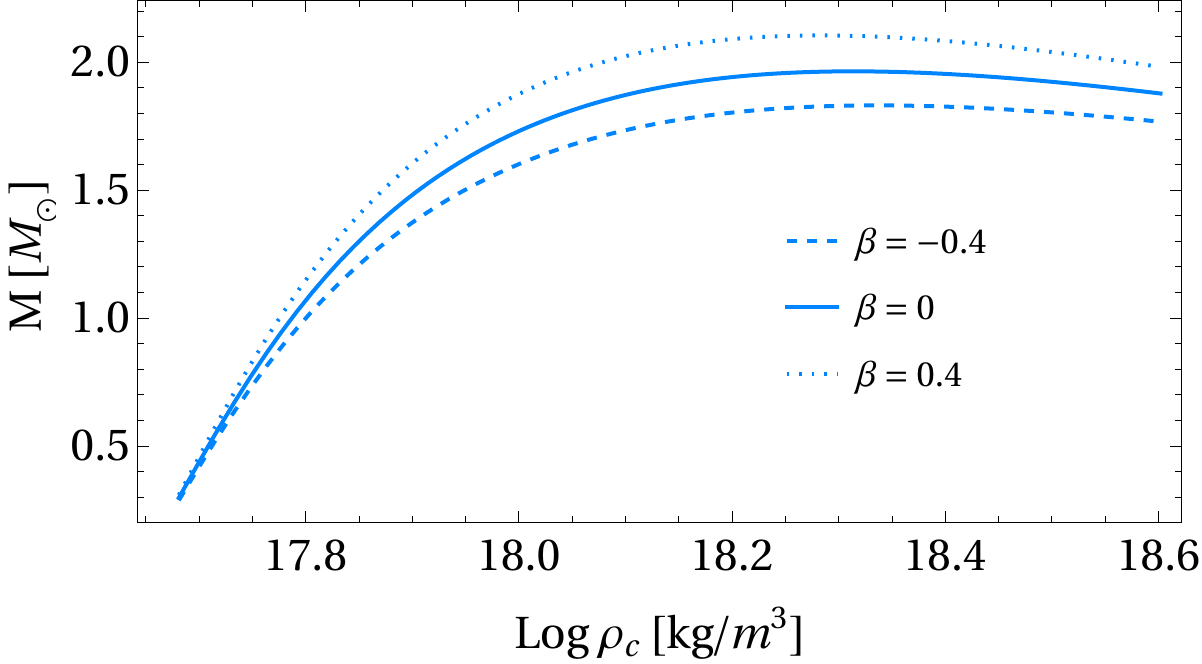}
  \includegraphics[width=8.6cm]{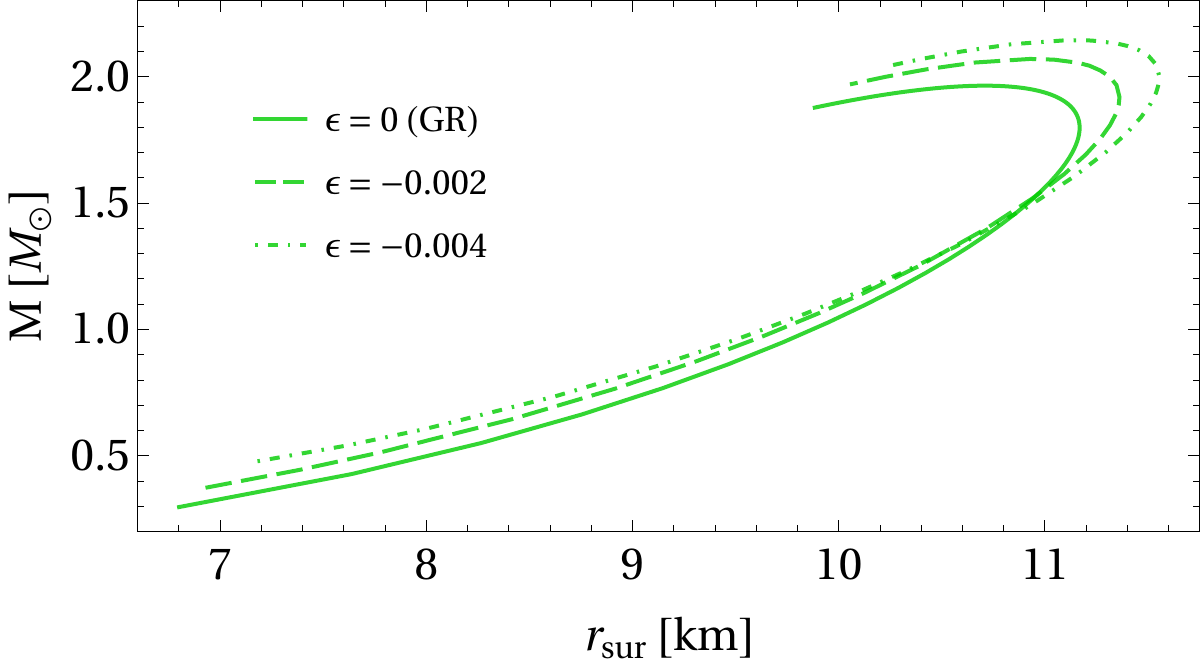}
  \includegraphics[width=8.6cm]{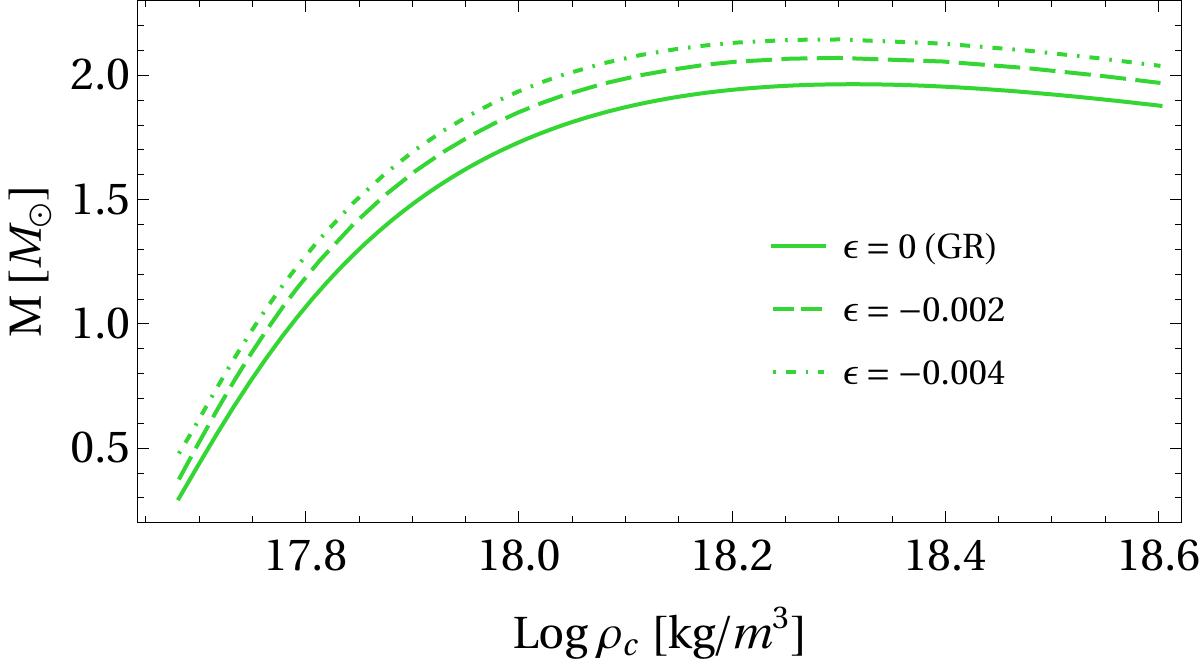}
  \includegraphics[width=8.6cm]{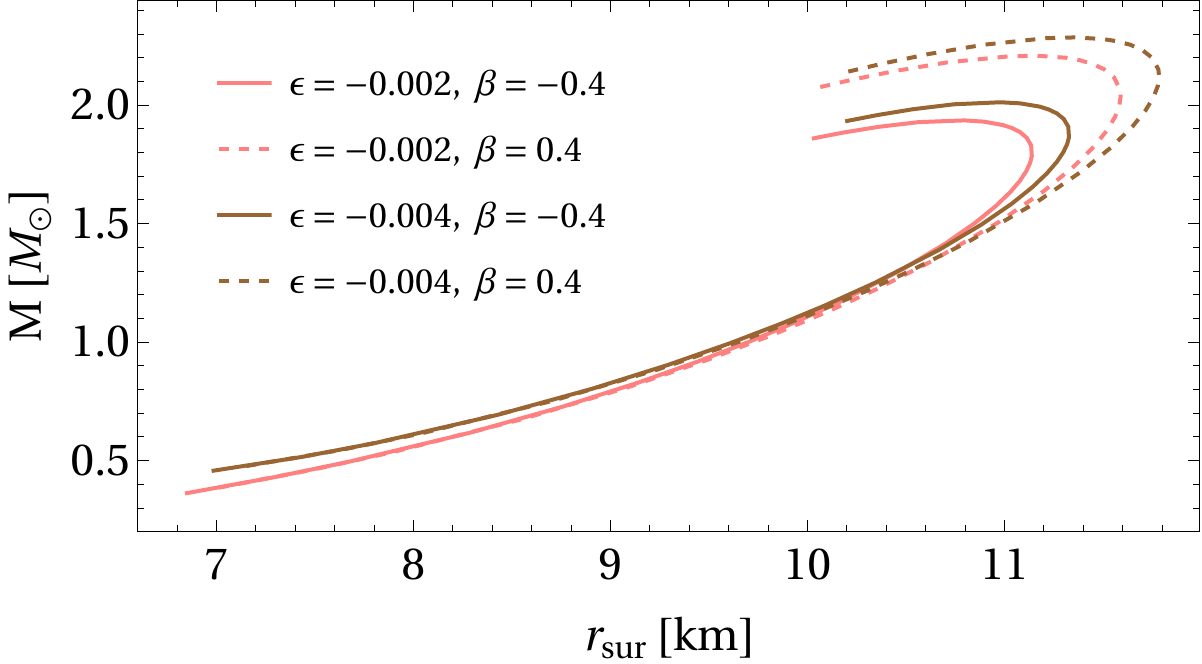}
  \includegraphics[width=8.6cm]{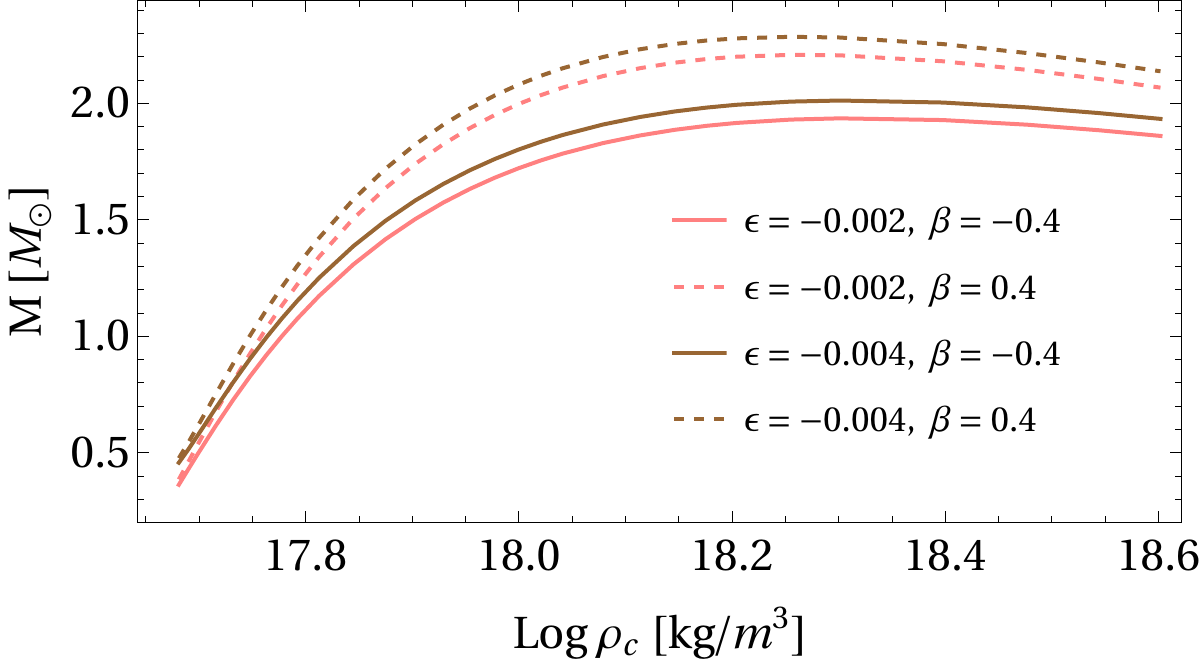}
  \caption{\label{figure4} Mass-radius diagrams (left column) and mass-central density relations (right column) for quark stars with MIT bag model EoS in GR ($\epsilon= 0$) and within the context of $f(R) = R^{1+\epsilon}$ gravity for some values of $\epsilon$. The blue curves in the upper panels represent anisotropic solutions for $\epsilon= 0$. Green lines in the middle panels correspond to isotropic solutions ($\beta =0$) for different values of $\epsilon$. Pink and brown curves in the lower plots correspond to $\beta \neq 0$ and $\epsilon\neq 0$. Note that some combinations of $\epsilon$ and $\beta$ are capable of generating maximum masses above $2M_\odot$. The curves exhibit a remarkable behaviour because the radius and mass increase as $\epsilon$ becomes more negative. Furthermore, the effect of anisotropy is more relevant in the high-mass region. }  
\end{figure*}

For different values of central energy density, we can obtain a family of anisotropic quark stars in $f(R)= R^{1+ \epsilon}$ gravity. Consequently, the mass-radius diagrams and mass-central energy density relations in such modified gravity framework (including the corresponding pure GR results for the anisotropic case by blue lines) are shown in Fig.~\ref{figure4}. As one can see, for $\beta$ fixed and $\epsilon$ varying, significant deviations from GR appear in both high-mass and low-mass regions. We observe that the $\epsilon R\ln R$ term in Eq.~(\ref{28}) allows for an increase in the maximum-mass values as $\epsilon$ becomes more negative. On the other hand, for $\beta$ varying and $\epsilon$ fixed, the total gravitational mass of quark stars undergoes very slight changes at low central densities and the most substantial changes occur at higher densities due to the anisotropic pressure. This qualitative behavior is similar to the general relativistic situation, see for instance the blue curves in the upper plots of Fig.~\ref{figure4}, where three values of the anisotropy parameter are particularly considered. Furthermore, we remark that negative values of $\beta$ lead to lower maximum masses as in Einstein gravity.

Unlike the Starobinsky model, where the quadratic term has a substantial effect only in the high-radius region for quark stars (see Ref.~\cite{Astashenok2015}), in the present work we observe that the logarithmic correction given by Eq.~(\ref{28}) has a significant impact on both high-radius and low-radius regions. Such deviations with respect to the pure GR framework can be visualized in the mass versus radius diagrams by green lines in Fig.~\ref{figure4}.

\begin{table}
\caption{\label{table1} 
Stellar configurations with central energy density $\rho_c = 1.0 \times 10^{18} \rm{kg}/\rm{m}^3$ and MIT bag model EoS (\ref{MITbagEoS}) in $f(R)= R^{1+ \epsilon}$ gravity for different values of $\epsilon$ and $\beta$. The total gravitational mass of the star as measured by a distant observer is denoted by $M$. The radial profile of the metric functions and Ricci scalar for the configuration corresponding to $\epsilon= -0.002$ and $\beta= 0.4$ is shown in Fig.~\ref{figure1}. Moreover, the radial behaviour of the pressures and mass function is shown in Fig.~\ref{figure2}. }
\begin{ruledtabular}
\begin{tabular}{l|ccc}
Parameters  &  $r_{\rm sur}$ [km]  &  $m_{\rm sur}$ [$M_\odot$]  &  $M$ [$M_\odot$]  \\
\colrule
$\epsilon= 0$ (GR), $\beta= 0$   &  11.151  &  1.729  &  1.729  \\
$\epsilon= -0.002$, $\beta= -0.4\ $  &  11.111  &  1.628  &  1.721  \\
$\epsilon= -0.002$, $\beta= 0$  &  11.344  &  1.759  &  1.851  \\
$\epsilon= -0.002$, $\beta= 0.4$  &  11.583  &  1.906  &  1.996  \\
$\epsilon= -0.004$, $\beta= -0.4$  &  11.299  &  1.656  &  1.801  \\
$\epsilon= -0.004$, $\beta= 0$  &  11.539  &  1.791  &  1.934  \\
$\epsilon= -0.004$, $\beta= 0.4$  &  11.779  &  1.936  &  2.080  \\
\end{tabular}
\end{ruledtabular}
\end{table}

\section{Conclusions}\label{Sec5}

In this work we have investigated the global physical properties of compact stars within the context of metric power-law $f(R)$ gravity. In particular, we considered the gravitational action $f(R) = R^{1+\epsilon}$ in order to explore small deviations from the usual Einstein-Hilbert action for $\vert \epsilon \vert \ll 1$. Under a non-perturbative formulation of the field equations, we have derive the modified TOV equations and studied the equilibrium structure of compact stars in the presence of anisotropy. We have adopted the anisotropy profile suggested by Horvat and collaborators \citep{Horvat2011}, where appears a dimensionless parameter $\beta$ which controls the degree of anisotropy within the compact star. In that sense, the progress made in this study was to extend previous works on isotropic compact stars in $f(R) = R^{1+\epsilon}$ gravity to include anisotropy. 

The mass-radius diagrams of anisotropic quark stars in terms of the free parameters $\epsilon$ and $\beta$ have been constructed for the MIT bag model EoS. We have found that the main effect of the parameter $\epsilon$ is to increase the total gravitational mass throughout the range of central energy densities. In addition, the relevant changes due to the anisotropic pressure emerge in the high-mass region (close to the maximum-mass point), while the variations are negligible in the low-mass region. Although this behavior is similar to that obtained in RG, it is also interesting to note that suitable combinations of the parameters $\epsilon$ and $\beta$ allow us to obtain masses above $2M_\odot$. This would make it possible to describe supermassive compact stars and hence have a better agreement between the theoretical calculations and the observational data.

\begin{acknowledgments}
JMZP acknowledges financial support from the PCI program of the Brazilian agency ``Conselho Nacional de Desenvolvimento Científico e Tecnológico''--CNPq. SBD thanks CNPq for partial financial support. This work has been done as a part of the Project INCT-Física Nuclear e Aplicações, Project number 464898/2014-5.
\end{acknowledgments}\


%

\end{document}